\documentclass{article}
\usepackage{spconf,amsmath,graphicx}
\usepackage[backend=bibtex,natbib=true, maxcitenames=1]{biblatex} %

\usepackage{xcolor}
\usepackage{subcaption}

\usepackage{balance}
\def\x{{\mathbf x}}
\def\L{{\cal L}}

\title{Training Keyword Spotters with Limited and Synthesized Speech Data}
\name{James Lin, Kevin Kilgour, Dominik Roblek, Matthew Sharifi}
\address{Google Research}
\begin{document}
\maketitle
\begin{abstract}
With the rise of low power speech-enabled devices, there is a growing demand to quickly produce models for recognizing arbitrary sets of keywords. As with many machine learning tasks, one of the most challenging parts in the model creation process is obtaining a sufficient amount of training data. In this paper, we explore the effectiveness of synthesized speech data in training small, spoken term detection models of around 400k parameters. Instead of training such models directly on the audio or low level features such as MFCCs, we use a pre-trained speech embedding model trained to extract useful features for keyword spotting models. Using this speech embedding, we show that a model which detects 10 keywords when trained on only synthetic speech is equivalent to a model trained on over 500 real examples. We also show that a model without our speech embeddings would need to be trained on over 4000 real examples to reach the same accuracy.
\end{abstract}

\section{Introduction}
\label{sec:intro}

With the proliferation of personal smart devices, there has been a corresponding increase in the use of voice control. Although it is possible to use full speech recognition systems that either run on the cloud or are optimized for on device usage, both of these approaches have disadvantages. Off-loading the computation to the cloud implies sending the continuous stream of audio recorded by the device to an external server, which raises many privacy concerns, increases the overall cost of the device and reduces its reliability.
And while on device speech recognition systems would alleviate these problems, they require an order of magnitude more power than keyword spotters designed to detect only the specific set of keywords relevant to the device's use. In cases where the commands issued to the smart device are more complex, the keyword spotter can be used as trigger for a full speech recognition system that is either run on the device or in the cloud.

A setup that allows for quick generation of models to recognize custom sets of keywords can allow device manufacturers to rapidly prototype and iterate during product development. It may also provide a mechanism for end users to customize the speech interfaces of their devices, giving them more control when interacting with their devices. %

In this paper, we investigate how many training examples we actually need to build a working keyword detector and examine two approaches of reducing the amount to a level that would more easily facilitate rapid prototyping. First we show how a speech embedding model trained to produce embeddings for keyword spotting can be used to reduce the number of examples required to train a keyword spotter on a new set of keywords from 35k down to under 2k.
We further demonstrate that replacing these 2k training examples with speech examples generated from a speech synthesizer only reduces the accuracy by about 2\% to 92.6\%.

\section{Related Work}
\label{sec:related_work}

There has been much related prior work in the areas of keyword spotting systems, augmentation with synthesized speech, transfer learning, and multi-task learning. Some of the work within the past few years in the area of keyword spotting include that of Chen et al \cite{Chen2014SmallfootprintKS} on using a DNN for recognizing the phrase ``okay google", the hybrid systems of TDNN-HMM \cite{Sun2017CompressedTD} and DNN-HMM \cite{raju2018data} used in Amazon's Alexa, and Sigitia et al's work for Apple's Siri \cite{Sigtia2018EfficientVT} (also of DNN-HMM architecture). There are also examples of using convolutional networks for keyword spotting. Couke et al \cite{Coucke_2019} employed the dilated convolutions of the WaveNet \cite{oord2016wavenet} architecture and showed results that were more robust to noise than both LSTM or regular CNN based models. The temporal convolution on the ResNet \cite{he2015deep} architecture of Choi et al \cite{choi2019temporal} is notable for achieving an accuracy of 96.6\% on the Speech Commands dataset \cite{warden2018speech} with only 350k parameters.

On the subject of transfer learning and multi-task learning, Sun et al \cite{Sun2017CompressedTD} use a full speech recognition system to initialize the weights of their keyword spotter as well as a parallel task where they share the bottom layers for the keyword spotter with the speech recognition system's acoustic model. Raju et al \cite{raju2018data} use a similar multi-target training setup for their keyword spotter.

The use of synthesized speech for data augmentation is also not new \cite{rygaard2015using, li2018training, rosenberg2019speech, Mimura2018LeveragingSS}. In particular, the work described in this paper utilizes a Tacotron 2 \cite{Shen_2018} based synthesizer to generate the synthetic speech examples, as is the case with the work by Li et al \cite{li2018training} and Rosenberg et al \cite{rosenberg2019speech}.

\section{Model Architecture}
\label{sec:model_architecture}
Our model is designed for deployment 
in an environment where both memory and compute power are very limited, such as on a digital signal processor (DSP). It runs on top of a low footprint feature extractor that provides a 32 dimensional log mel feature vector covering the frequency range from 60 Hz to 3800 Hz, quantized to 8 bits every 10 ms. Under the assumption that much of the work required to classify a small set of keywords is independent of the actual keywords, we split the model into two parts: an \textit{embedding model} that contains 5 convolutional blocks ($\sim$330k weights) and a \textit{head model} that contains a single convolutional block along with a classification block ($\sim$55k weights).

Each detection requires an input context of 198 feature vectors (approximately 2s), which is provided by a low memory footprint feature extractor. This is followed by 6 convolutional blocks and a classification block. Each convolutional block consists of 5 layers: a 1x3 convolution, a 3x1 convolution, a maxpool layer, a 1x3 convolution, and a 3x1 convolution. The final block only uses 3x1 convolutions as the frequency dimension is 1 after the previous layers. Starting with 24 channels in the first block, we increase the number of channels by 24 in each new block until a maximum of 96 is reached.
The classification block can be either a maxpool layer followed by a 1x1 convolution if a continuous stream of  predictions is required, or a maxpool layer followed by a fully connected layer. The results reported Section~\ref{sec:results} are based on evaluations on single words where the fully connected layer variant is used.

\subsection{Embedding Model}
\label{sec:model_architecture:embedding_model}
\begin{figure}[t]
    \centering
    \includegraphics[width=0.8\columnwidth]{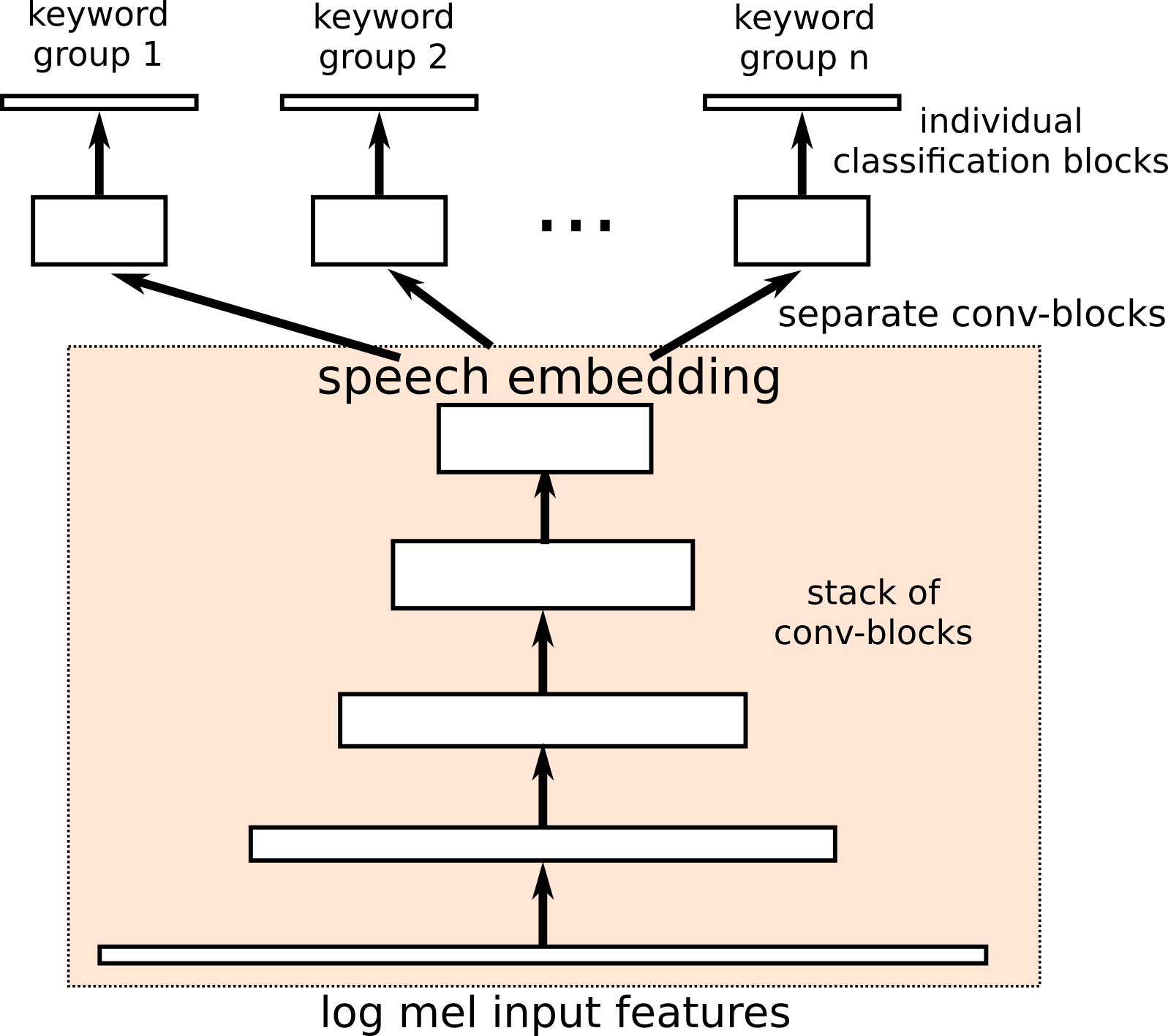}
    \caption{\textit{Training setup of the embedding model (light orange block). Multiple head models with different groups of target keywords are trained in parallel on top of the shared embedding model which learns an embedding that should be useful for any arbitrary group of keywords. After training, the head models are discarded. \vspace{-0.3cm}}}
    \label{fig:model_train}
\end{figure}

The embedding model converts a stream of audio into a stream of 96-dimensional feature vectors, one every 80 ms. To ensure that the resulting embedding is useful for arbitrary sets of keywords, we took 5000 keywords (most of them are actually 2-3 words long) and split them into random groups of 40 keywords. The resulting 125 groups of keywords are used to train 125 keyword spotting models with shared weights for the embedding model part (see Figure~\ref{fig:model_train}). We used roughly 200 million 2-second audio clips from YouTube for training, of which 100 million contained the target keywords and the other 100 million were used as non-target examples. This embedding model is available on Tensorflow Hub for public use. The embedding model is trained using TensorFlow \cite{abadi2016tensorflow} on 20 GPUs for 2 days and is available on TensorFlow Hub (\url{https://tfhub.dev/google/speech_embedding/1}) for reuse. 
 
\subsection{Head Model}
We include the embedding model in our graph as a TensorFlow Hub module with fine-tuning switched off. A head model can be trained on single GPU in under 30 minutes. On small data sets, this can be much faster; with models trained on 100k examples converging in 180 seconds, and models trained on 1000 examples converging in around 30-40 seconds. This head model architecture scales well with the number of keywords, requiring only an extra 96 weights per new target keyword. A fully connected model on the same input would require 1536 weights per new target keyword.

\section{Experimental Setup}
\label{sec:experiment_setup}
The data used in our experiments can be categorized into two types: real speech data and synthesized speech data. In addition to the YouTube data used to train the embedding model, we also use the Speech Commands dataset \cite{warden2018speech} as a source of our real speech data. The speech commands dataset contains 80k short utterances of 35 different words, 10 of which are considered target words. It is not used at all in the training of the embedding model; the embedding model is trained only on the YouTube data. All reported accuracy numbers are on the Speech Commands evaluation set.

\subsection{Baseline and the Effects of Limited Data}
A baseline is established by training and testing on the whole speech commands dataset.
For both the full model and the head model on top of a pre-trained embedding model, we evaluate the effects of training on a limited amount of data by creating training sets of various sizes. These training sets are created by selecting between 1 and 1000 random examples from the Speech Commands training set for each of the 35 speech command words. Since the Speech Commands dataset contains 35 words, this results in models trained  on 35 to 35000 examples. To take into consideration the variability in the selections, for each choice of training set size, we perform the selection and training 20 times and compute the mean and standard deviation.

\subsection{Synthesized Speech Data}
We synthesize each of the 35 speech command words using 92 different voices\footnote{We originally synthesized 100 examples per word, but later realized that 8 voices were duplicates and removed them.} with a text-to-speech system based on Tacotron 2 \cite{Shen_2018}, with considerations for fast synthesis that could be reasonably run on device. The quality of these examples is similar to what can be generated using the Google Cloud Text-to-Speech API \cite{cloudAPI}.

We examine two use cases for synthetic data, either as a replacement for real data or as additional data that could be used to supplement existing real data. To test the first use case, we begin by training models on only synthetic speech and gradually replace random subsets of various sizes with real examples. For the second case, we start with the synthesized speech data and add sets of real speech data randomly selected from the Speech Commands training set. For both cases, again we perform each data selection and model training 20 times and compute their mean and standard deviation.

In addition, we also experimented with creating additional synthetic speech examples from the directly synthesized speech examples via traditional data augmentation techniques. We tried both pitch- and tempo-shifts, as well as adding room reverberation effects and white noise. Unfortunately, the preliminary results were disappointing, as we saw no significant improvements in accuracy when compared to using only the directly synthesized speech data. This parallels the findings of \cite{qi2018low} where their augmentation of the small amount of data available for a new class failed to improve its classifier performance. We also suspect this may be due to the embedding model already having learned to deal with these distortions.

\begin{table}[t]
    \centering
    \caption{\textit{Performance of the full and head models trained on synthetic data, an equivalent amount of real data, and the full speech commands training data set.}}
     \label{tab:model_accuracy}
    \begin{tabular}{l|c|c|c}
        Training data           &  Size   &   Full model          &   Head model         \\\hline
       speech commands         &  80k    &   $       97.4 \% $   &   $      97.7 \% $   \\
        synthetic data          &  3220   &   $       56.7 \% $   &   $      92.6 \% $   \\
        equivalent real data    &  3220   &   $       88.7 \% $   &   $      95.3 \% $   
    \end{tabular}
\end{table}

\begin{figure}[t]
    \centering
    \includegraphics[width=0.99\columnwidth]{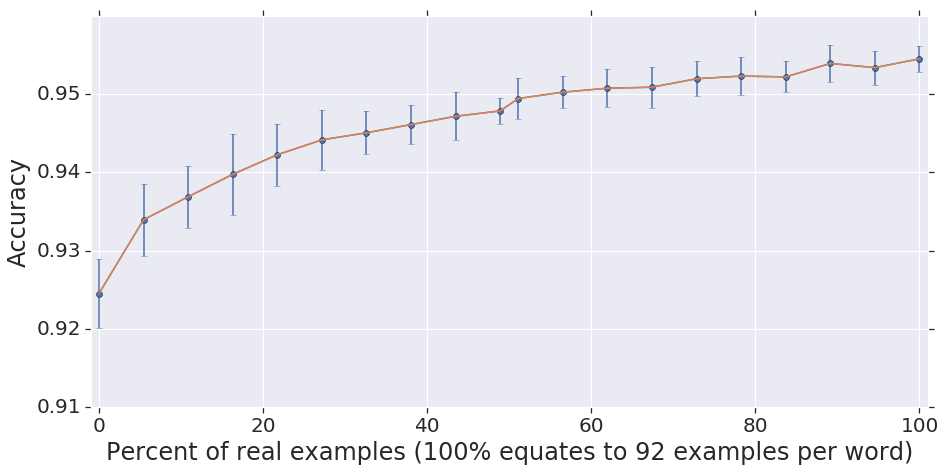}
    \caption{\textit{Performance with different proportions of real vs synthetic speech examples. \vspace{-0.3cm}}}
    \label{fig:real_replaced_syn}
\end{figure}

\begin{figure*}[t]
    \centering
    \includegraphics[width=0.9\textwidth]{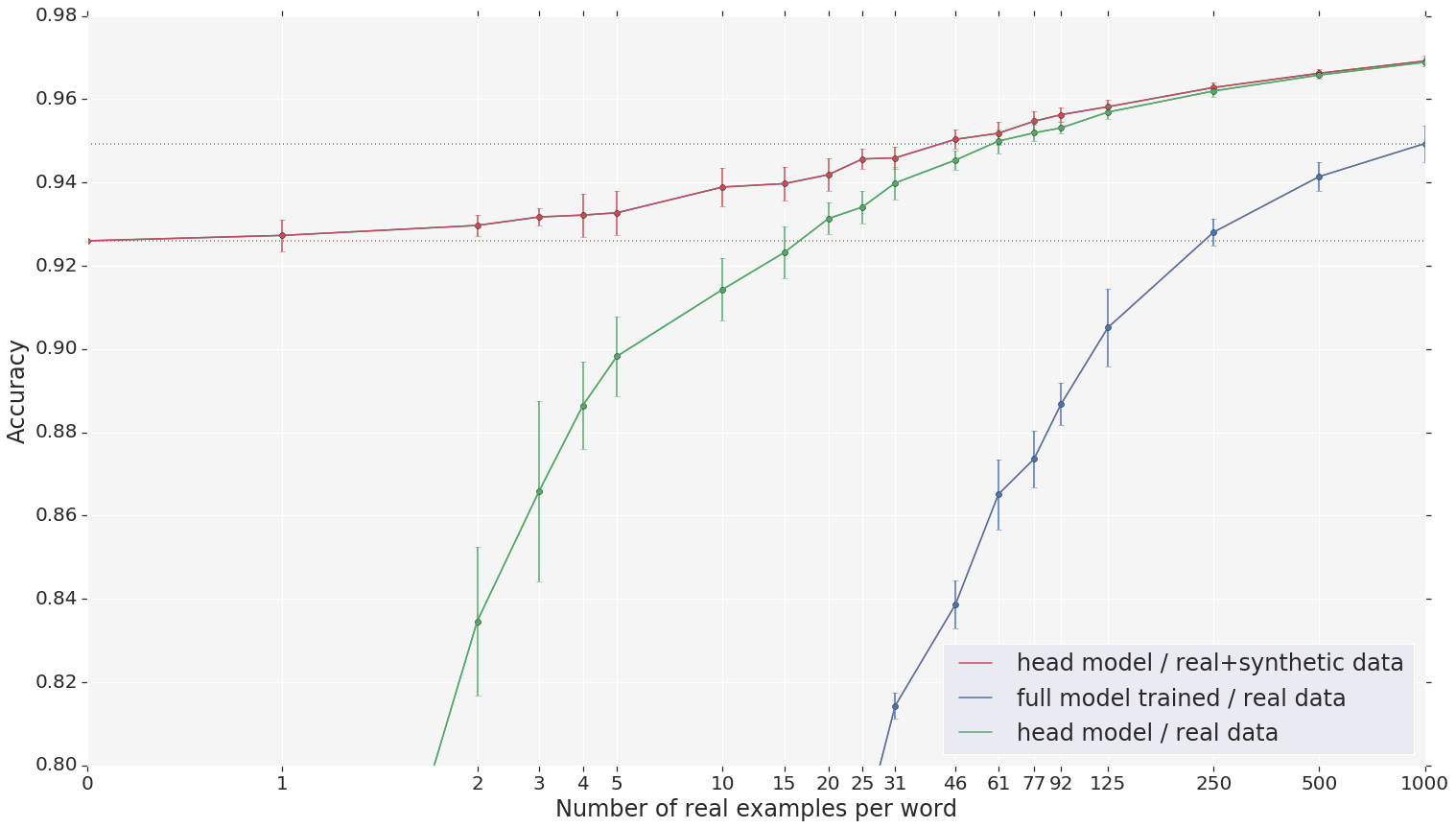}
    \vspace{-0.1cm}
    \caption{\textit{Performance of different amounts of real speech examples, with and without augmentation using synthetic speech examples, optionally based on the speech embedding model. The standard deviation over 20 runs is shown for each data point. \vspace{-0.3cm}}}
    \label{fig:real_plus_syn}
\end{figure*}

\section{Results}
\label{sec:results}
For use as baselines, we trained both a full model and a head model using the full speech commands data. In Table~\ref{tab:model_accuracy} we compare the models trained using only our generated set of 3220 synthetic examples and models trained on an equivalent number of real examples.
Training the full model on only the synthetic examples results in an accuracy of 56.7\%. Using the same number of real examples, the accuracy increases to 88.7\%. With the full speech commands dataset, we are able to obtain an accuracy of 97.4\%.

Using our speech embeddings improves the performance of the head model trained on the full speech commands dataset only slightly, to 97.7\%. We can, however, see significant gains on the head models trained on small numbers of examples. Using 3k real examples, we see the accuracy increase from 88.7\% to 95.3\%, which is equivalent to a 58\% relative decrease in error. Performance of the model trained on only synthetic speech jumps to 92.6\% with a relative error rate reduction of 83\%.

\subsection{Replacing Real Data}
These last two results can be seen as the two data points at the extremes of the possibilities between using only real speech data and using only synthetic data to train the head model. We explore in Figure~\ref{fig:real_replaced_syn} some intermediate points on this continuum, where we use a certain percentage of real examples with the remaining being synthetic. The plot shows the mean and variance of the results based on 20 random selections per percentage-combination. Overall the error bars are quite large and indicate a lot of variability between different random selections of training data. Despite this variability, we see a clear general tendency that by simply replacing a small number of synthetic examples with real examples, we already bridge a large part of the gap between the two extremes. After replacing about 50\% of the data with real data, the curve begins to saturate as the accuracy differences of the models have already been reduced by almost 90\%.

\subsection{Augmenting Real Data}
We explored the use of synthetic data to augment the real data that already exists, or to reduce the amount of real data that needs to be collected. In real applications, this approach is more realistic, as one would almost certainly use all the synthetic speech examples that one has, and go about collecting increasing amounts of real speech examples.

Figure~\ref{fig:real_plus_syn} shows these results. At the right of the plot we can see that when we have 1000 real examples per word (35k total), using the embedding model improves the accuracy noticeably from 94.8\% to 96.8\%. The inclusion of the synthetic speech data does not result in any further improvements. Decreasing the number of real examples to 125 per word reduces the accuracy of the full model by 4.5\% (absolute) to 90.3\%, while the head model, at an accuracy of 95.8\%, only loses 1\% (absolute). This is also the first point where adding the synthetic data produces a measurable improvement of 0.1\%. Reducing the number of real examples further causes the performance of the full model to rapidly decrease.

As we decrease the number of real examples, before dropping below 50 examples per word, the performance of the head model trained on only real examples is better than or on par with the full model trained on 1000 examples per word. Head models that are also trained on synthetic data remain at least as accurate as the full model trained on 1000 examples per word until the number of real examples drops below about 25. This is also when we see a significant difference between the models trained on only real examples and models trained on those augmented with synthetic data.

When the training data has 10 real examples per word, the difference from having the synthetic data has grown to 2.5\% (absolute) and reaches 3\% when training data has only 5 real examples per word. This is also the point where the accuracy of the head model trained without synthetic data drops below the accuracy of a full model trained on 125 real examples per word. The accuracy of head models augmented with synthetic data is consistently higher than the accuracy of a full model trained on 125 examples per word. This is true even when the number real examples being augmented is 0, so that the head model is in fact trained on only synthetic data.

\vspace{-0.2cm}

\section{Conclusion}
\label{sec:conclusion}

In this paper, we demonstrate the effectiveness of using synthesized speech data for training keyword spotting systems in the context of a speech embedding model.
While synthesized speech does not provide enough useful information to train a full keyword spotter, it can be used to train a head model on top of a well trained speech embedding model. The accuracy of such a head model is equivalent to the accuracy of a head model using 50 real examples per word, or a full model trained on over 4000 real examples per word. Our small head model can be trained in a few minutes on a modern GPU. In future work we plan to investigate how to efficiently train the model on a small device with limited compute resources.
\vspace{-0.2cm}

\section{Acknowledgments}
We would like to thank Marco Tagliasacchi, Dick Lyon, Andrew Rosenberg and Bhuvana Ramabhadran for their help and feedback.

\vfill\pagebreak

\label{sec:refs}

\balance

\printbibliography

\end{document}